\documentclass[12pt,amstex,amssymb]{article}
\usepackage{amssymb}
\usepackage{amsmath}
\usepackage{graphicx}
\usepackage{color}
\usepackage{hyperref}

\newcommand{\R}{{\mathbb R}}
\newcommand{\Z}{{\mathbb Z}}
\renewcommand{\S}{{\bf S}}

\newcommand{\G}{{\mathcal G}}
\newcommand{\B}{{\frak B}}
\newcommand{\hkq}{/\!\!/\!\!/}
\newcommand{\M}{{\cal M}}
\renewcommand{\H}{{\mathbb H}}
\newcommand{\C}{{\mathbb C}}

\let\oldsqrt\sqrt
\def\sqrt{\mathpalette\DHLhksqrt}
\def\DHLhksqrt#1#2{%
\setbox0=\hbox{$#1\oldsqrt{#2\,}$}\dimen0=\ht0
\advance\dimen0-0.2\ht0
\setbox2=\hbox{\vrule height\ht0 depth -\dimen0}%
{\box0\lower0.4pt\box2}}

\title{Moduli Spaces of Instantons\\ on the Taub-NUT Space}
\author{
Sergey A. Cherkis\thanks{E-mail: cherkis@maths.tcd.ie}\\
\it School of Mathematics and\\
\it Hamilton Mathematics Institute,\\
\it Trinity College, Dublin, Ireland}

\begin{document}
\begin{titlepage}

\renewcommand{\thepage}{ }
\date{}

\maketitle
\abstract{We present ADHM-Nahm data for instantons on the Taub-NUT space and encode these data in terms of Bow Diagrams.  We study the moduli spaces of the instantons and present these spaces as finite hyperk\"ahler quotients.  As an example, we find an explicit expression for the metric on the moduli space of one $SU(2)$ instanton.  

We motivate our construction by identifying a corresponding string theory brane configuration. By  following string theory dualities we are led to supersymmetric gauge theories with impurities.} 

\vspace{-5.7in}

\parbox{\linewidth}
{\small\hfill \shortstack{TCDMATH 08-05\\ \hfill HMI 08-01}}

\end{titlepage}

\tableofcontents

\section{Introduction}
The celebrated construction of Atiyah, Drinfeld, Hitchin, and Manin \cite{Atiyah:1978ri} provided a description of all instantons of $\R^4$ in terms of algebraic data.  It has been generalized in a number of ways.  Werner Nahm in \cite{Nahm} discovered the construction of calorons, i.e. instantons on $\R^3\times \S^1,$ as well as that of magnetic monopoles, in terms of solutions of  an integrable system of Ordinary Differential Equations.  Kronheimer and Nakajima \cite{KN} constructed instantons on ALE spaces in terms of algebraic data organized into a quiver diagram.  
Nekrasov and Schwarz \cite{Nekrasov:1998ss} modified the original ADHM construction to obtain instantons on noncommutative $\R^4.$ Here, we present data describing instantons on the Taub-NUT and multi-Taub-NUT spaces \cite{TN}, which combines all elements of the above mentioned constructions.

We motivate our construction by string theory analysis of a Chalmers-Hanany-Witten \cite{Chalmers:1996xh, Hanany:1996ie} brane configuration that is T-dual to the brane realization of instantons on a Taub-NUT space.  This analysis is akin to the string theory derivation of the original ADHM and Kronheimer-Nakajima constructions for A-type ALE spaces by Douglas and Moore \cite{DM, Douglas}.  D-brane analysis leading to the Kronheimer-Nakajima construction on a general ALE space appeared in \cite{Johnson:1996py}.  

The Kronheimer-Nakajima construction was used to obtain some explicit instanton solutions on the Eguchi-Hanson space in \cite{Bianchi:1996zj} and \cite{Bianchi:1995xd}.
For the case of calorons, the explicit solutions were found in \cite{Kraan:1998kp} and \cite{Lee:1998bb}. 

A detailed transform from the data we present here to the gauge field on the Taub-NUT, as well as an explicit solution of one instanton on the Taub-NUT, will appear in \cite{InstOnTN}.

Let us emphasize that the AHDM and Nahm constructions produce instantons on flat backgrounds $\R^4$ and $\R^3\times\S^1$. Kronheimer-Nakajima construction does leads to instantons in nontrivial geometric backgrounds of ALE spaces, however, it is based on the fact that any ALE space is a deformation of an orbifold $\R^4/\Gamma$ of flat space.  Here, we study Yang-Mills instantons on essentially curved ALF spaces which do not possess any useful flat limit.  

There was a number of attempts at construction of instantons on the Taub-NUT space. 
Some isolated solutions were found in \cite{Pope:1978kj, Kim:2000mg} and particular families of solutions appear in \cite{Etesi:2003ei, Etesi:2002cc, Etesi:2001fb}. We claim that the construction presented here produces all solutions for generic boundary conditions. We explore the geometry of the moduli space of solutions and present as an illustration the explicit metric on the moduli space of charge one instanton.

\section{Self-dual Connections on Taub-NUT}
A Taub-NUT space is described by the following metric
\begin{equation}\label{Eq:TN}
ds^2=\frac{1}{4}\left[ \left(l+\frac{1}{|\vec{r}|}\right)d\vec{r}^{\,2}+\frac{\left(d\tau+\omega\right)^2}{\left(l+\frac{1}{|\vec{r}|}\right)} \right],
\end{equation}
where $\tau\sim\tau+4\pi$ is the periodic coordinate on the Taub-NUT, $\omega$ is a one-form $\omega=\vec{\omega}\cdot d\vec{r}\,$  such that $d\omega=*_3 d\frac{1}{|\vec{r}|}$ (here $*_3$ is the Hodge star operation for a flat three-dimansional space parameterized by $\vec{r}$).

An instanton on this space is a Hermitian connection $A$ with the curvature $F=dA-i A\wedge A$ such that it has finite action  and 
the curvature form is self-dual: 
\begin{equation}
F=* F.
\end{equation}
Here $*$ denotes the Hodge dual operation for the Taub-NUT space (\ref{Eq:TN}).  There is a number of topological invariants associated to such a connection.  The action is given by the integral of the Chern character:
\begin{equation}
S=-\frac{1}{8\pi^2}\int_{\rm Taub-NUT} {\rm tr} F\wedge F,
\end{equation}
and monopole charges are defined in the following way.  Since the action is finite, the curvature tends to zero as we approach infinity, thus the connection tends to a flat one on the squashed three-sphere $\S^3_r: \left\{ |\vec{r}|=r\right\}$ as $r\rightarrow\infty.$  This three-sphere is Hopf-fibered over $\S^2_r,$ with the fiber $S^1$ parameterized by $\tau.$ The fiber has a finite size at infinity.  Since the limiting connection is flat, the monodromy of the connection along the fiber $\S^1$ has eigenvalues independent of the point on the base $\S^2_r.$ Let the limiting values of the eigenvalues be  $\exp(2\pi i \lambda/l)$ and $\exp(-2\pi i \lambda/l),$ with $0\leq\lambda\leq l/2,$ as $r\rightarrow\infty.$  Generically $\lambda\neq 0$ and thus the bundle over $\S^2_r$ splits into eigen line bundles $L_+\rightarrow\S^2_r$ and $L_-\rightarrow\S^2_r,$ with degrees $d_+$ and $d_-.$  We define the monopole charge by $m=|d_+-d_-|.$ 

In intuitive physics terms, at infinity the gauge field becomes independent of $\tau,$ and after the dimensional reduction, it looks like a monopole field. The charge $m,$ defined above, is the charge of this monopole.  To be more precise, let us denote the limiting $\tau$-independent connection along the $\tau$-circle by $A_\tau d\tau$ and the other three horizontal components of the connection by $A_j,$ so that $A=A_j dx^j+A_\tau d\tau.$ Dimensionally reducing along the finite $\tau$ circle, as in \cite{Kronheimer}, we obtain the Higgs field  $\Phi=\left(\l+\frac{1}{|\vec{r}|}\right) A_\tau$ and the gauge field $A'_j=A_j-\omega_j A_\tau$ in $\R^3.$  If $F'$ denotes the three-dimensional curvature of $A',$ the pair $(\Phi, A')$ satisfies the Bogomolny equation $F=*_3 [D,\Phi].$  Thus at infinity the pair $(\Phi, A')$ behaves as a monopole and $m$ is its charge.

As demonstrated in \cite{Kronheimer} the case of nonzero $m$ and vanishing instanton number reduces to the study of singular monopoles.  Nahm data for singular monopoles was identified in \cite{Cherkis:1997aa} and further explored in \cite{Cherkis:1998hi, Cherkis:1998xc}.   Explicit singular monopole solutions were constructed in \cite{Cherkis:2007qa, Cherkis:2007jm}.
In this paper we focus on the  pure instanton case, i.e we put $m=0.$

\section{Instanton Data as a Bow Diagram}\label{Sec:Bow}
We consider an $SU(2)$ instanton of  zero monopole charge and instanton number $N$ with maximal symmetry breaking, that is, with eigenvalues of the monodromy matrix at infinity $\exp(\pm 2\pi i \lambda/l)$ with $l>\lambda>0.$  Each such instanton configuration is determined uniquely by the data we describe below.  This data can be organized into the diagram in Figure~\ref{Fig:Bow}.
\begin{figure}[htbp]
\label{Fig:Bow}
   \centering
   \includegraphics[width=6cm]{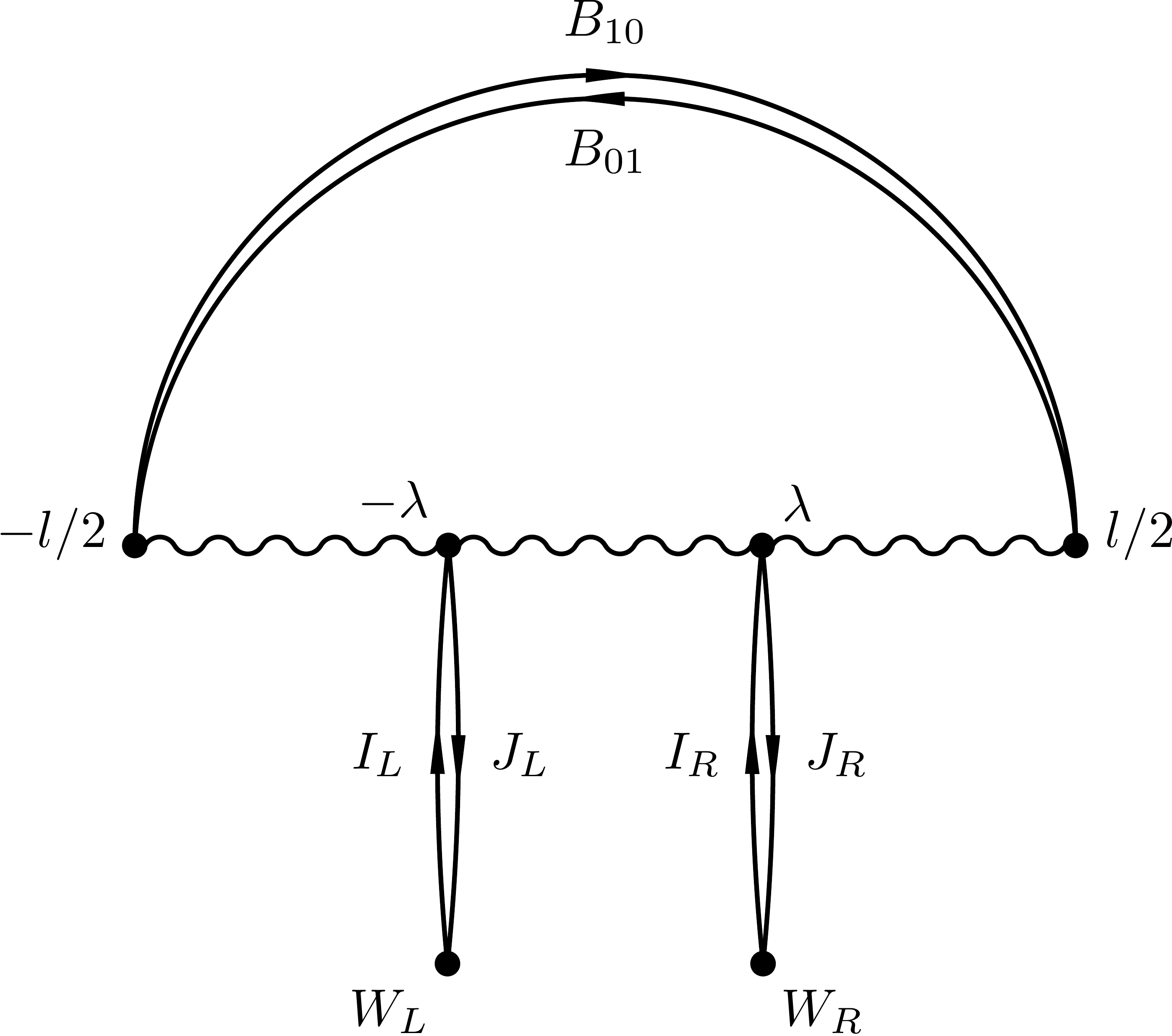} 
   \caption{Bow diagram corresponding to $SU(2)$ Instantons on Taub-NUT}
   \label{Fig:Bow}
\end{figure}
\linebreak
In the limit $l\rightarrow 0$ the Taub-NUT space degenerates to flat $\R^4$ and the above diagram becomes the ADHM quiver diagram for instantons on $\R^4.$   We shall refer to diagrams such as the one in Figure \ref{Fig:Bow} as Bow Diagrams. 

Each interval represented by a wavy line connecting two dots corresponds to  $U(N)$ Nahm data. In the diagram above there are three such intervals $[-l/2.-\lambda], [-\lambda, \lambda],$ and $ [\lambda, l/2];$ we shall refer to these as `Left', `Middle,' and `Right' intervals respectively and parameterize each by a coordinate $s$ taking value in these ranges.  We denote the lengths of these intervals by $d_L=l/2-\lambda, d_M=2\lambda,$ and $d_R=l/2-\lambda.$ Nahm data consists of a quadruplet $(T_0(s), T_1(s), T_2(s), T_3(s))$ of $s$-dependent Hermitian $N\times N$ matrices continuous on each interval and satisfying the Nahm equations
\begin{align}\label{Nahm}
i \frac{d}{ds}T_1+[T_0, T_1]-[T_2, T_3]=0,\nonumber\\
i \frac{d}{ds}T_2+[T_0, T_2]-[T_3, T_1]=0,\\
i \frac{d}{ds}T_3+[T_0, T_3]-[T_1, T_2]=0.\nonumber
\end{align}
Geometrically, one can think of a Hermitian vector  bundle $E$ of complex dimension $N$ over each interval with a connection $\frac{d}{ds}-i T_0$ and Hermitian Higgs fields (i.e. endomorphisms of the bundle $E$)  $T_1, T_2,$ and $T_3.$ We use subscripts and superscripts $L, M,$ and $R$ to specify the interval to which the Nahm data belong.  The dots on the wavy line represent the  fibers of this bundle $E_{-l/2}, E_{-\lambda}, E_\lambda,$ and $E_{l/2}$ over the points $s=-l, -\lambda, \lambda,$ and $l.$ External dots represent one dimensional vector spaces $W_L$ and $W_R.$ Each arrow connects two dots and represents a map from the space at its tail to the space at its head. For example, in any given  trivialization $B_{10}$  is represented by an $N\times N$ matrix corresponding to a map from $E_{-l/2}$ to $E_{l/2}.$

This data transforms under a unitary gauge transformation $g(s)\in U(N)$ as
\begin{equation}\label{Eq:Transformation}
g(s): \left(\begin{array}{c}
T_0(s)\\
T_j(s) \\
 \\
B_{01}\\
B_{10}\\
 \\
I_L \\
J_L\\
 \\
I_R\\
J_R
         \end{array}
	     \right)\mapsto
	     \left(\begin{array}{c}
g^{-1}(s) T_0 g(s)-i g^{-1}(s)\frac{d}{ds}g(s)\\
g^{-1}(s) T_j g(s)\\
 \\
g^{-1}(-l/2) B_{01} g(l/2)\\
g^{-1}(l/2) B_{10} g(-l/2)\\
 \\
g^{-1}(-\lambda) I_L \\
J_L g(-\lambda)\\
 \\
g^{-1}(\lambda) I_R \\
J_R g(\lambda)
    	    \end{array}
	     \right)
\end{equation}

Besides the Nahm Eqs. (\ref{Nahm}), these data satisfy certain  conditions at $s=\pm l$ and $\pm\lambda;$  to write these in a compact form we introduce an auxiliary twistorial variable $\zeta$ and a combination
\begin{equation}
A=T_1+i T_2+2 \zeta T_3-\zeta^2(T_1-i T_2),
\end{equation} 
for each interval.  Then the following conditions are satisfied for all values~of~$\zeta:$
\begin{align}\label{Eq:jumps1}
A^R(l/2)&=-(B_{10}+\zeta B_{01}^\dagger)(B_{01}-\zeta B_{10}^\dagger)\\
\label{Eq:jumps2}
A^L(-l/2)&=(B_{01}+\zeta B_{10}^\dagger)(B_{10}-\zeta B_{01}^\dagger)\\
\label{Eq:jumps3}
A^R(\lambda)-i A^M(\lambda)&=-(I_R+\zeta J_R^\dagger)(J_R-\zeta I_R^\dagger)\\
\label{Eq:jumps4}
A^M(-\lambda)-i A^L(-\lambda)&=-(I_L+\zeta J_L^\dagger)(J_L-\zeta I_L^\dagger)
\end{align}
Let us point out that the matching conditions (\ref{Eq:jumps3}) and (\ref{Eq:jumps4}) are identical to those of a monopole Nahm data \cite{Hurtubise:1989qy}.

We claim that the set of bow data $(T_0, T_1, T_2, T_3, I_L, J_L, I_R, J_R, B_{01}, B_{10})$ satisfying Eqs.~(\ref{Nahm}, \ref{Eq:jumps1}-\ref{Eq:jumps4}) determines the instanton configuration on the Taub-NUT space up to a gauge transformation.  The equivalence classes of the bow data up to the transformation (\ref{Eq:Transformation}) are in one-to-one correspondence with such instantons.  Moreover, this correspondence is an isometry of the corresponding hyperk\"ahler moduli spaces.  For the case of the Nahm transform for monopoles Hitchin proved that it is one-to-one in \cite{Hitchin:1983ay}, and the proof of the isometry statement was given by Nakajima in \cite{Nakajima:1990zx}.

\section{Structure of Moduli Spaces}
The moduli space of instantons on the Taub-NUT is hyperk\"ahler; moreover, the corresponding bow data presents it as an infinite hyperk\"ahler quotient of linear spaces, since the Nahm equations (\ref{Nahm}) and the matching conditions of Eqs.~(\ref{Eq:jumps1}-\ref{Eq:jumps4}) can be viewed as moment maps with respect to the gauge group action of Eq.~(\ref{Eq:Transformation}).  The moduli space can also be viewed as a finite hyperk\"ahler quotient of a product of linear spaces and a number of $T^*U^{\C}(N)=T^*Gl(N).$  The hyperk\"ahler structure on the the cotangent bundle to  a group was studied in \cite{DancerSwann}, where this space emerges as a moduli space of Nahm data with regular boundary conditions on an interval.
It carries a natural triholomorphic action of $G\times G$ with the first and second factors corresponding to the value of the gauge transformation on one and the other end  of the interval.  In terms of these building blocks the moduli space $\M_{N, \lambda; l}$ of charge $N$ instantons on the Taub-NUT space  (\ref{Eq:TN}) is given by the following hyperk\"ahler quotient:
\begin{equation}\label{Eq:finite}
T^*G^\C_{d_L}\times\H^N\times T^*G^\C_{d_M}\times \H^N \times T^*G^\C_{d_R}\times \H^{N^2}\hkq G_{-l/2}\times G_{-\lambda}\times G_{\lambda}\times G_{l/2},
\end{equation}
where $\H^N\approx\C^N\times\C^N\ni(I^\dagger, J),  \H^{N^2}\approx\C^{N^2}\times\C^{N^2}\ni(B_{01}^\dagger, B_{10}),$  each group $G$ is a $U(N)$ and 
\begin{itemize}
\item $G_{-l/2}$ acts on $T^*G^\C_{d_L}$ on the left and on $\H^{N^2}$ on the right, 
\item $G_{-\lambda}$ acts on the first $\H^N$ factor, acting on $T^*G^\C_{d_L}$ on the right and on $T^*G^\C_{d_M}$ on the left, 
\item $G_{\lambda}$ acts on the second $\H^N$ factor, acting on $T^*G^\C_{d_M}$ on the right and on $T^*G^\C_{d_R}$ on the left, 
\item $G_{l/2}$ acts on $T^*G^\C_{d_R}$ on the right and on $\H^{N^2}$ on the left. 
\end{itemize}
This allows one to construct more or less explicitly the twistor space of $\M_{N, \lambda; l}.$ Let us now explain how this finite quotient construction arises from our bow diagram description.

Let us use the language of the hyperk\"ahler reduction to give geometric meaning to the equations of Section \ref{Sec:Bow}.  Given the bow data  one views the space of all such unconstrained data 
\begin{equation}
\B=\left\{(T_0(s), T_1(s), T_2(s), T_3(s), I_L, J_L, I_R, J_R, B_{01}, B_{10})\right\},\end{equation}
as a direct product of infinite-dimensional  vector spaces of unconstrained Nahm quadruplets $(T_0(s), \vec{T}(s))$ for each interval and the linear spaces of $(B_{01}, B_{10}),$ $(I_L, J_L),$ and $(I_R, J_R).$ The metric on the space of Nahm quadruplets on each interval is  given by 
\begin{equation}
ds^2 =\int {\rm tr} \left(\delta T_0 \delta T_0^\dagger+\delta T_1 \delta T_1^\dagger+\delta T_2 \delta T_2^\dagger+\delta T_3 \delta T_3^\dagger\right)ds,
\end{equation}
and the natural metrics on the rest of the data are 
\begin{equation}
\delta I_L^\dagger \delta I_L+ \delta J_L \delta J_L^\dagger, \quad  {\rm tr} \left(\delta B_{01} \delta B_{01}^\dagger+ \delta B_{10} \delta B_{10}^\dagger\right), \quad  \delta I_R^\dagger \delta I_R+ \delta J_R \delta J_R^\dagger.
\end{equation}
There are three natural complex structures on these spaces coming from identifying $T_0, T_1, T_2, T_3$ as components of a quaternion and from the identification of the spaces parameterized by $B$'s and the pairs $(I,J)$ with quaternions. These complex structures are spelled explicitly in Section~\ref{Sec:OneInst}.

The transformations (\ref{Eq:Transformation}) leave the metric and the complex structures invariant. Let $\G$ denotes the group of all $U(N)$ gauge transformations on $[-l/2, l/2].$ Our claim is that the moduli space of instantons $\M_{N, \lambda; l}$ is isometric to the  hyperk\"ahler quotient of $\B$ by $\G:$
\begin{equation}
\M_{N, \lambda; l}=\B\hkq\G
\end{equation}
$\B\hkq\G$ is exactly the space of equivalence classes under the gauge transformation (\ref{Eq:Transformation}) of bow data satisfying the moment map Eqs.~(\ref{Nahm},\,\ref{Eq:jumps1}-\ref{Eq:jumps4}).
This is the infinite hyperk\"ahler quotient of linear spaces. Now we compare this formula to Eq.(\ref{Eq:finite}). Let $\G_0$ denote the subgroup of $\G$ consisting of all gauge transformations $g(s)$ that equal to identity at the marked points (i.e with  $g(-l/2)=g(-\lambda)=g(\lambda)=g(l/2)=1),$ then 
\begin{equation}
\G/\G_0=G_{-l/2}\times G_{-\lambda}\times G_{\lambda}\times G_{l/2}.
\end{equation}
Thus, we can perform the above reduction $\B\hkq\G$ in two steps, first performing the quotient with respect to $\G_0$ and then with respect to $\G/\G_0.$  The moment maps of the $\G_0$ action are exactly the left-hand-sides of the Nahm Eqs.({\ref{Nahm}).  Taking the zero level of the moment map and dividing by the group action we are left with $T^*G^\C_d$ on each interval. The second step of the hyperk\"ahler reduction amounts to formula (\ref{Eq:finite}).

\subsection{Algebraic Description}
Let us give a more algebraic description of this space.  Selecting a particular complex structure leads to a description of the Nahm data as a connection $D$ on the vector bundle and its endomorphism $T$ (none of these are restricted to be  Hermitian).  For example, for one of the complex structures, $D=\frac{\partial}{\partial s}-i T_0-T_3$ and $T=T_1+i T_2.$    With the complex structure selected, we can combine the three Nahm equations into one complex and one real.
  The complex Nahm equation within the interval simply reads $[D, T]=0.$ Now, we introduce the parallel transport $H$ from one end of the interval to another. Let $H(s)$ be covariantly constant with respect to $D$ (i.e. $D H(s)=0$) such that the value of $H(s)$ at the left end of the coresponding interval equals identity.  Then we denote the value of $H(s)$ at the right end of the interval by $H.$  This yields  us natural complex coordinates $(H,T)$ on $T^*G^\C,$ which is the moduli space of all the regular Nahm data satisfying the Nahm equations.  

As in \cite{Donaldson:1985id}, in a given complex structure the moduli space $\M_{N, \lambda; l}=\{\vec{\mu}=\vec{0}\}/\G$ is equivalent as a complex variety to $\{\mu^\C=0\}/ \G^\C.$ The latter is given by 
\begin{equation}
\left\{(T_L, H_L, I_L, J_L, T_M, H_M, I_R, J_R, T_R, H_R, B_{01}, B_{10})\right\},
\end{equation} 
where 
\begin{align}
T_{L,M,R}&\in gl(N,\C),&  I_{L,R}&\in {\rm Hom}(\C,\C^N),&   B_{01}&\in {\rm Hom}(\C^N, \C^N),\\
H_{L,M,R}&\in Gl(N, \C),& J_{L,R}&\in {\rm Hom}(\C^N, \C),& B_{10}&\in {\rm Hom}(\C^N, \C^N),
\end{align} 
satisfying the complex moment map conditions
\begin{align}
T_R- H_M^{-1} T_M H_M&=I_R J_R,& H_R^{-1}T_R H_R&=B_{10} B_{01},\\
T_M-H_L^{-1}T_L H_L&=I_L J_L,& T_L&=B_{01} B_{10},
\end{align}
modulo the gauge equivalence
\begin{equation}
\left(
\begin{array}{cc}
T_L,& H_L\\ I_L,& J_L\\ T_M,& H_M\\ I_R,& J_R\\ T_R,& H_R\\ B_{01},& B_{10}
\end{array}
\right)\sim 
\left(
\begin{array}{cc}
h^{-1}_{-l/2} T_L h_{-l/2},& h^{-1}_{-l/2} H_L h_{-\lambda}\\ 
h_{-\lambda}^{-1} I_L,& J_L h_{-\lambda}\\ 
h_{-\lambda}^{-1}T_M h_{-\lambda},& h_{-\lambda}^{-1} H_M h_\lambda\\ 
h_\lambda^{-1} I_R,& J_R h_\lambda\\ 
h_\lambda^{-1} T_R h_\lambda,& h_\lambda^{-1} H_R h_{l/2}\\ 
h_{l/2}^{-1} B_{01} h_{-l/2},& h_{-l/2}^{-1}B_{10} h_{l/2}
\end{array}
\right),
\end{equation}
here $h_{-l/2}, h_{-\lambda}, h_\lambda,$ and $h_{l/2}\in Gl(N, \C).$

\subsection{Comparison with AHDM Construction}
We can use the gauge transformations given by $h_{-\lambda}, h_\lambda,$ and $h_{l/2}$ to put $H_L=H_M=H_R=1$, then $T_L=B_{01}B_{10},\, T_M=B_{01}B_{10}-I_R J_R,\, T_R=B_{10}B_{01},$ and the only nontrivial remaining relation is 
\begin{equation}\label{Eq:ADHM}
B_{10}B_{01}-B_{01}B_{10}=I_L J_L+I_R J_R,
\end{equation}
with the gauge equivalence given by the remaining group with the action of an element $h=h_{-l/2}$
\begin{equation}\label{Eq:equiv}
h: \left(\begin{array}{cc}B_{01},& B_{10}\\ I_L,& J_L\\ I_R,& J_R\end{array}\right)\mapsto
\left(\begin{array}{cc}h^{-1}B_{01}h,& h^{-1}B_{10} h\\ h^{-1}I_L,& J_L h\\ h^{-1}I_R,& J_R h\end{array}\right).
\end{equation}
Eq.(\ref{Eq:ADHM}) with the equivalence (\ref{Eq:equiv}) is exactly the ADHM condition for instantons on the flat space. 

This establishes the isomorphism of the moduli space of instantons on a Taub-NUT space and the moduli space of unframed instantons on $\R^4$ as complex varieties.  We would like to emphasize, however, that even though in any given complex structure these two moduli spaces are isomorphic, their twistor spaces and metrics differ.

\section{Some Generalizations}
\subsection{$\mathbf U(\mathbf{n})$ Instantons}
$N$ instantons on the Taub-NUT space with the gauge group $U(n)$ and with maximal symmetry breaking at infinity are given in terms of the bow diagram in Figure~\ref{Fig:UnTN}.

\begin{figure}[h!]
\label{Fig:Jelly}
   \centering
   \includegraphics[width=6cm]{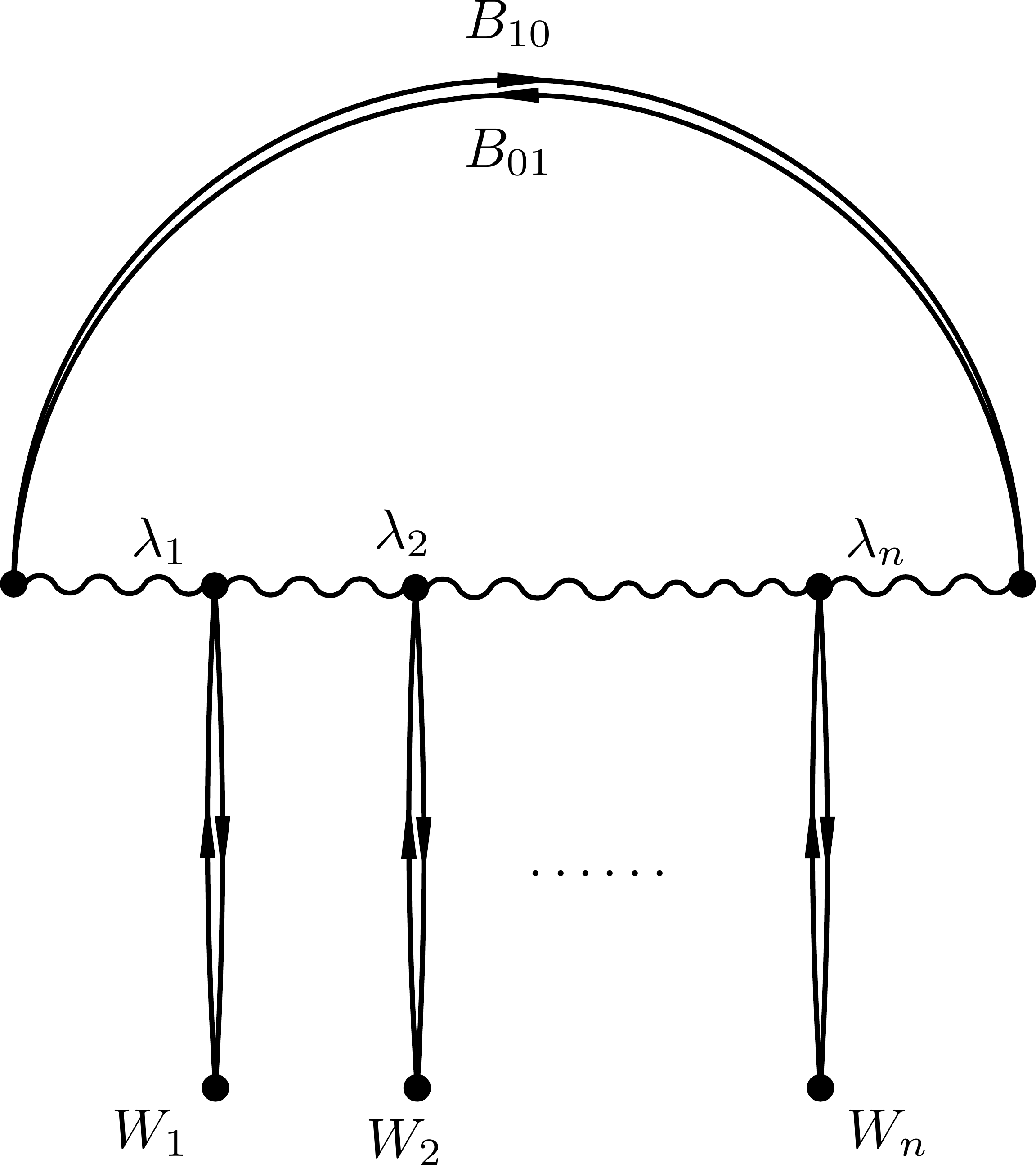} 
   \caption{A Bow corresponding to $SU(n)$ Instantons on Taub-NUT}
   \label{Fig:UnTN}
\end{figure}

Here $\exp(2\pi i\lambda_1/l), \exp(2\pi i\lambda_2/l), \ldots, \exp(2\pi i\lambda_n/l)$ are the eigenvalues of the monodromy at infinity and, if all of them are distinct, the auxiliary spaces $W_1, W_2, \ldots, W_n$ are one-dimensional. The moduli space is again given by a finite hyperk\"ahler quotient of a product of linear spaces and $n+1$ copies of $T^*Gl(N).$

\subsection{Instantons on multi-Taub-NUT}
Given the bow data for $SU(2)$ instantons on the Taub-NUT it is relatively easy to obtain the bow data on a general $A_k$ ALF space, which is a $(k+1)$-centered multi-Taub-NUT space. In order to achieve this, similarly to \cite{KN}, one can consider the quotient of the Taub-NUT by the cyclic group $\Z_{k+1}$ rotating the Taub-NUT circle $\tau\rightarrow\tau+\frac{4\pi}{k+1}.$  If we are to obtain $N$ instantons on $(k+1)$-centered degenerate Taub-NUT, we should consider rank $N(k+1)$  ADHM data for its covering Taub-NUT space.  In other words, the bundle $E$ of section \ref{Sec:Bow} is rank $N(k+1)$.  The equivariance conditions one should impose are
\begin{equation}\label{Eq:Equivariance}
g(s): \left(\begin{array}{c}
T_0(s)\\
T_j(s) \\
 \\
B_{01}\\
B_{10}\\
 \\
I_L \\
J_L\\
 \\
I_R\\
J_R
         \end{array}
	     \right)\mapsto
	     \left(\begin{array}{c}
U^{-1} T_0 U\\
U^{-1} T_j U\\
 \\
U^{-1} B_{01} U e^{\frac{2\pi i}{k+1}}\\
e^{-\frac{2\pi i}{k+1}}U^{-1} B_{10}U\\
 \\
U^{-1} I_L \mu_L \\
\mu_L^{-1}J_L U\\
 \\
U^{-1} I_R \mu_R\\
\mu_R J_R U
    	    \end{array}
	     \right).
\end{equation}
Here the multiplication by $e^{\pm\frac{2\pi i}{k+1}}$ rotates the circle of the Taub-NUT, and the factors $\mu_L$ and $\mu_R$ with $|\mu_L|=|\mu_R|=1$ are the linear transformations of the auxiliary spaces $W_L$ and $W_R.$ The factors $\mu_L$ and $\mu_R$ correspond to the accompanying gauge transformation of the gauge fields on the Taub-NUT. 
$U={\mathbb I}_{N\times N}\otimes C_{k+1},$ with $C_{k+1}$ being a permutation matrix permuting the $(k+1)$ blocks. We can choose it to be a clock matrix $C_{k+1}={\rm diag} (1, e^{\frac{2\pi i}{k+1}}, e^{2\frac{2\pi i}{k+1}},\ldots, e^{k\frac{2\pi i}{k+1}}).$   In order to satisfy the equivariance conditions $\mu_L$ and $\mu_R$ have to be some integer powers of the $(k+1)$st root of unity, say $\mu_L=e^{\frac{2\pi i}{k+1} p_L}$ and $\mu_R=e^{\frac{2\pi i}{k+1} p_R}.$

The equivariant bow data thus consists of the block diagonal $T_0$ and $T_j,$
\begin{equation}
B_{01}=\left(\begin{array}{cccccc}
\mbox{\scriptsize 0}& \mbox{\scriptsize 0}&\mbox{\scriptsize 0}&\mbox{\scriptsize $\ldots$}&\mbox{\scriptsize $B_{01}^{(k)}$}\\
\mbox{\scriptsize $B_{01}^{(1)}$}&\mbox{\scriptsize 0}&\mbox{\scriptsize 0}&\mbox{\scriptsize $\ldots$}&\mbox{\scriptsize 0}\\
\mbox{\scriptsize 0}&\mbox{\scriptsize $B_{01}^{(2)}$}&\mbox{\scriptsize 0}&\mbox{\scriptsize $\ldots$}&\mbox{\scriptsize 0}\\
\mbox{\scriptsize $\vdots$}&\mbox{\scriptsize $\vdots$}&\mbox{\scriptsize $\vdots$}&\mbox{\scriptsize $\ddots$}&\mbox{\scriptsize $\vdots$}\\
\mbox{\scriptsize 0}&\mbox{\scriptsize 0}&\mbox{\scriptsize 0}&\mbox{\scriptsize $\ldots$}&\mbox{\scriptsize 0}
\end{array}\right),\
B_{10}=\left(\begin{array}{cccccc}
\mbox{\scriptsize 0}&\mbox{\scriptsize $B_{10}^{(1)}$} &\mbox{\scriptsize 0}&\mbox{\scriptsize $\ldots$}&\mbox{\scriptsize 0}\\
\mbox{\scriptsize 0}&\mbox{\scriptsize 0}&\mbox{\scriptsize $B_{10}^{(2)}$} &\mbox{\scriptsize $\ldots$}&\mbox{\scriptsize 0}\\
\mbox{\scriptsize 0}&\mbox{\scriptsize 0}&\mbox{\scriptsize 0}&\mbox{\scriptsize $\ldots$}&\mbox{\scriptsize 0}\\
\mbox{\scriptsize $\vdots$}&\mbox{\scriptsize $\vdots$}&\mbox{\scriptsize $\vdots$}&\mbox{\scriptsize $\ddots$}&\mbox{\scriptsize $\vdots$}\\
\mbox{\scriptsize $B_{10}^{(k)}$}&\mbox{\scriptsize 0}&\mbox{\scriptsize 0}&\mbox{\scriptsize $\ldots$}&\mbox{\scriptsize 0}
\end{array}\right),
\end{equation}
and $I_L$ and $J_L$ are such that they have only $p_L$-th block nonvanishing; similarly $I_R$ and $J_R$ have only the $p_R$-th block nonvanishing.
The remaining gauge transformations that respect the equivariance conditions are block diagonal.

These data are naturally arranged into a bow diagram with $k+1$ wavy lines connected cyclically by the components $B_{01}^{(j)}$ and 
$B_{10}^{(j)}$ and two auxiliary spaces $W_L$ and $W_R$ connected to some points on the $p_L$-th and $p_R$-th wavy lines.  

If we are to start with $U(n)$ instantons on the Taub-NUT (instead of the $SU(2)$ instantons discussed so far in this section) the only difference would be
to have $n$ auxiliary spaces and $n$ (instead of two) pairs of maps $(I_l, J_l).$ The maps $I_l$ and $J_l$ map between these auxiliary spaces and some fibers above $n$ points on some of the wavy intervals. 

Thus, the bow data for the $U(n)$ instantons in a nondegenerate multi-Taub-NUT background is given by Figure~\ref{Fig:UnmultiTN}. 
The positions of the dots on the wavy lines are given by the eigenvalues of the monodromy at infinity while the differences 
of the lengths of the wavy lines are determined by the fluxes of the $B$ field on the multi-Taub-NUT.  Just as in the ALE case \cite{KN}, for a generic (nondegenerate) multi-Taub-NUT, the moment maps no longer vanish, but are given by the multi-Taub-NUT resolution parameters.
\begin{figure}[htbp]
   \centering
   \includegraphics[width=6cm]{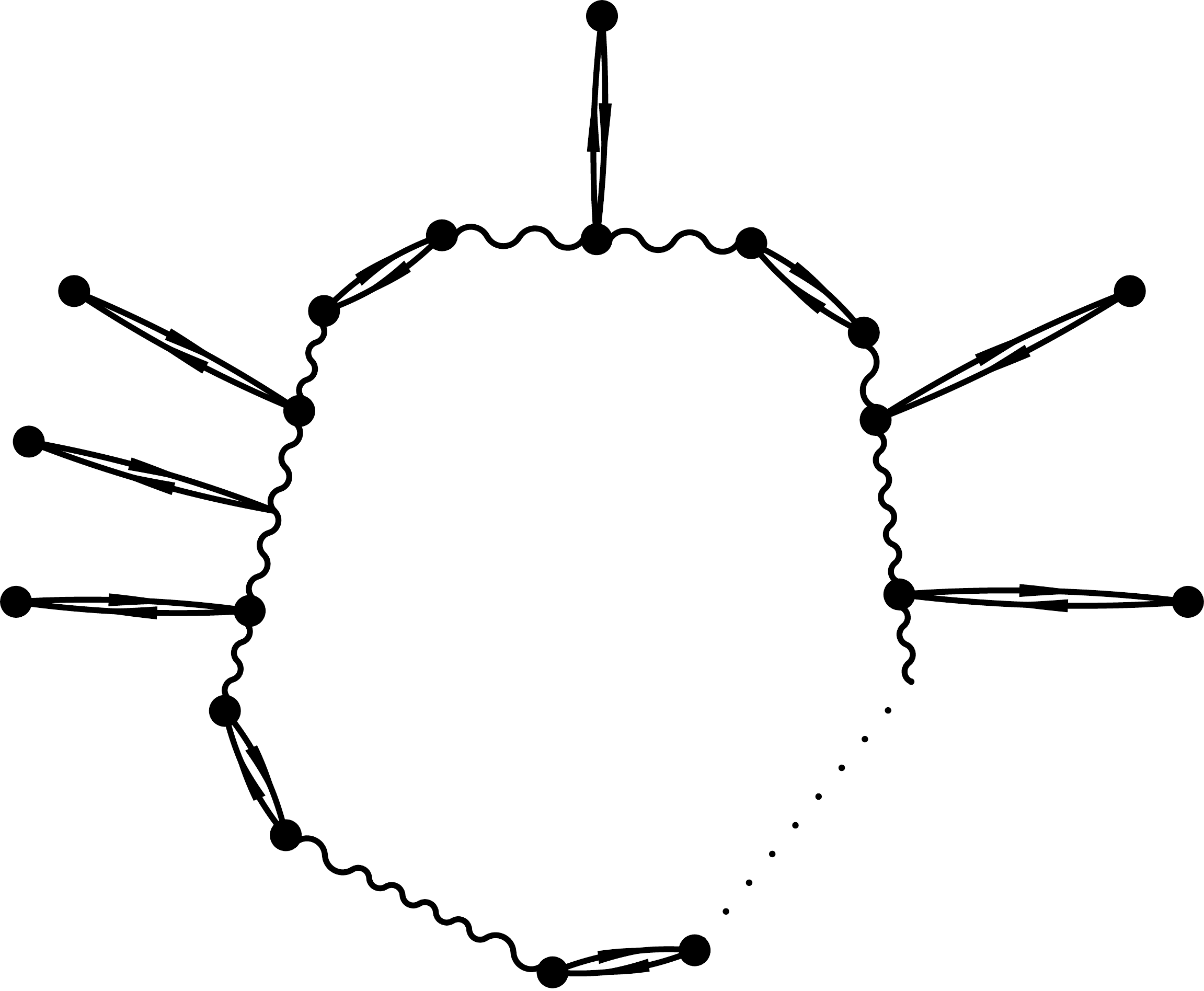} 
      \caption{A Bow corresponding to $U(n)$ Instantons on multi-Taub-NUT}
   \label{Fig:UnmultiTN}
\end{figure}

\section{Brane Configuration Analysis}\label{Sec:Branes}
We begin with a setup similar to that of Douglas and Moore \cite{DM} by realizing 
a $U(n)$ charge $N$ instanton configuration on a Taub-NUT space in Type IIA string theory.  Namely, consider Typer IIA string theory on a direct product of the Taub-NUT space and the six-dimensional Minkowski space-time. Let the coordinates $0,1,2,7,8,9$ be in the Minkowski space and the remaining coordinates $3,4,5$ and $6$ in the Taub-NUT space with the sixth coordinate being the periodic coordinate of the circle of the Taub-NUT.  Now introduce $n$ D6-branes wrapping the Taub-NUT space with the world-volumes in the $0,1,2,3,4,5,6$ directions, and $N$ D2-branes positioned at points on the Taub-NUT space with world-volumes in the $0,1,2$ directions.  In the effective world-volume theory on the D6-brane, this configuration is described by $N$ instantons in the $U(n)$ gauge group on the Taub-NUT.

Performing the T-duality along the periodic Taub-NUT direction $6,$ we obtain a Chalmers-Hanahy-Witten configuration \cite{Chalmers:1996xh, Hanany:1996ie} of $n$ D5-branes, $N$ D3-branes and one NS5-brane in a flat ten-dimensional Minkowski space with  direction $6$ periodic.  The exact orientation of the branes is specified in Table \ref{Table}.  
\begin{table}[htdp]
\begin{center}
\begin{tabular}{c|cccccccccc}
 &0&1&2&3&4&5&6&7&8&9\\
\hline
D5&$\times$ & $\times$& $\times$& $\times$& $\times$& $\times$& & & &  \\
\textcolor{red}{D3}& $\textcolor{red}{\times}$& $\textcolor{red}{\times}$& $\textcolor{red}{\times}$& & & & $\textcolor{red}{\times}$& & &  \\
\textcolor{blue}{NS5}& $\textcolor{blue}{\times}$& $\textcolor{blue}{\times}$& $\textcolor{blue}{\times}$& & & & & $\textcolor{blue}{\times}$& $\textcolor{blue}{\times}$&  $\textcolor{blue}{\times}$\\
\\
\hline
{\small Vector
}& {\small A}$_0$&  {\small A}$_1$& {\small A}$_2$& & & & & {\small Y}$_1$&  {\small Y}$_2$& {\small Y}$_3$\\
{\small Adjoint\,Hyper}& & & &{\small Re\,H$_2$}& {\small Im\,H$_2$}& {\small Re\,H$_1$}&{\small Im\,H$_1$}& & & 
\end{tabular}
\end{center}
\caption{Brane configuration and bosonic bulk matter content of the impurity theory on the D3-branes.}
\label{Table}
\end{table}%

Now, that we have identified the relevant brane configuration, we would like to describe the theory on the world-volume of the D3-branes.  In the absence of the five-branes this would be a maximally supersymmetric Yang-Mill theory with the gauge group $U(N).$  As indicated in Table \ref{Table} the Vector multiplet consists of the three-dimensional gauge field with components $A_0, A_1, A_2$ and the Higgs fields $Y_1, Y_2, Y_3,$ corresponding to the transverse directions $7,8,9$ to the D3-branes as well as two Majorana fermions. The Adjoint Hupermultiplet consists of one Dirac fermion, three Hermitian Higgs fields and one Hermitian connection.  We combine these Higgs fields and this connection into two complex fields $H_1$ and $H_2,$ so that ${\rm Re} H_2, {\rm Im} H_2, {\rm Re} H_1,$  are the Higgs fields  corresponding to the transverse directions $3,4,5,$ while ${\rm Re} H_1$ is the gauge field connection in direction $6.$ 
  
The presence of the five-branes breaks the maximal ${\cal N}=4$ supersymmetry to ${\cal N}=2,\, d=4$ and introduces additional degrees of freedom in the effective world-volume theory on the D3-brane.  Even though the sigma model analysis of this configuration is difficult due to the presence of both Ramond and Neveu-Schwarz charges,  let us give some local arguments for the inhomogeneities we are about to introduce. 

\subsection{D3-D5 Intersection}
Let us focus first on one of the D5-branes with a number of D3-branes ending on it on the left and the same number ending on the right as in the leftmost diagram in Figure \ref{Fig:D3D5}.  Assembling the D3-branes together, we can end up in the middle configuration of Figure \ref{Fig:D3D5}.  Now, we can move the D3-branes in the direction transverse to the D5, thus ending with the configuration on the right diagram of Figure \ref{Fig:D3D5}.  This brane configuration contains a massive string state corresponding to the lowest excitation of the string connecting the D3's and the D5's. Its  mass is proportional to the distance between the D3-branes and the  D5-brane and it is clearly in the fundamental representation of the gauge group on the D3-branes. 

Running this argument backwards, moving the D3's so that they intersect the D5 renders this fundamental multiplet $f$ massless, and separating the two parts of the D3-brane amounts to giving it a vacuum expectation value.
\begin{figure}[h]
  \begin{center}
   \includegraphics[width=13.7cm]{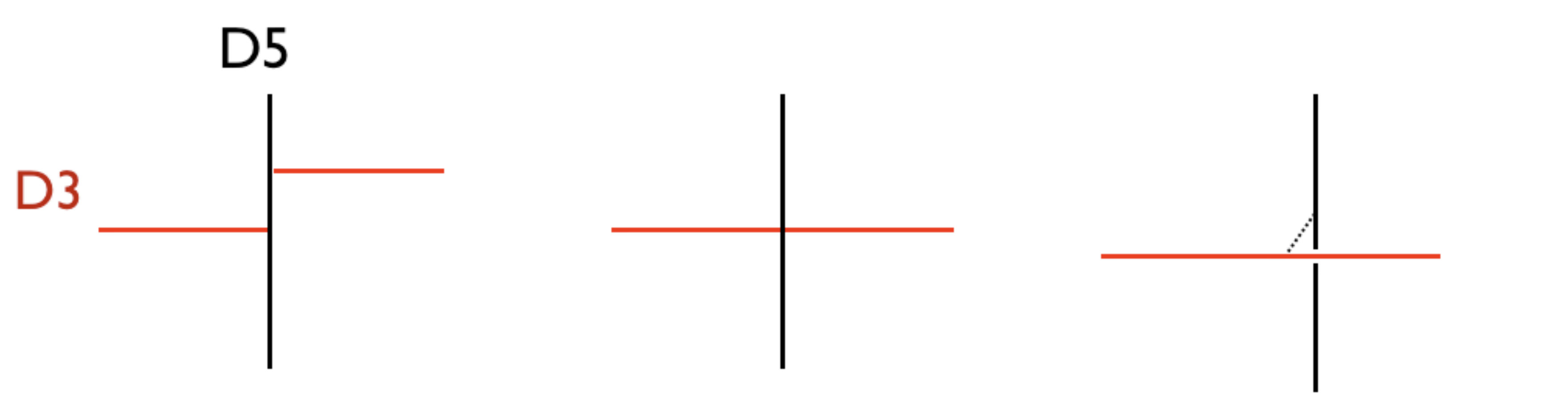} 
   \caption{Fundamental Multiplet}
   \label{Fig:D3D5}
   \end{center}
\end{figure}

\subsection{D3-NS5 Intersection}
Here we focus on the local configuration with the D3-branes in the vicinity of an NS5-brane as on the left diagram in Figure \ref{Fig:D3NS5}.  Moving the D3-branes towards the NS5-brane, so that they intersect as in the middle diagram, we can now separate the left and the right parts of the D3's along the NS5-brane, as in the rightmost diagram of Figure \ref{Fig:D3NS5}.  This configuration has a string state corresponding to the lowest excitation of a string stretching between the left and right D3-brane ends.  This state is in the bifundamental of the gauge group on the D3-brane and its mass is proportional to the distance between the two D3 ends.  

Again, running the arguments in reverse, we move the left and right D3-branes' ends together, so that the bifundamental multiplet $B$ becomes massless.  Now moving the D3-brane off the NS5-brane amounts to giving this multiplet some vacuum expectation value.
\begin{figure}[h]
  \begin{center}
   \includegraphics[width=13.7cm]{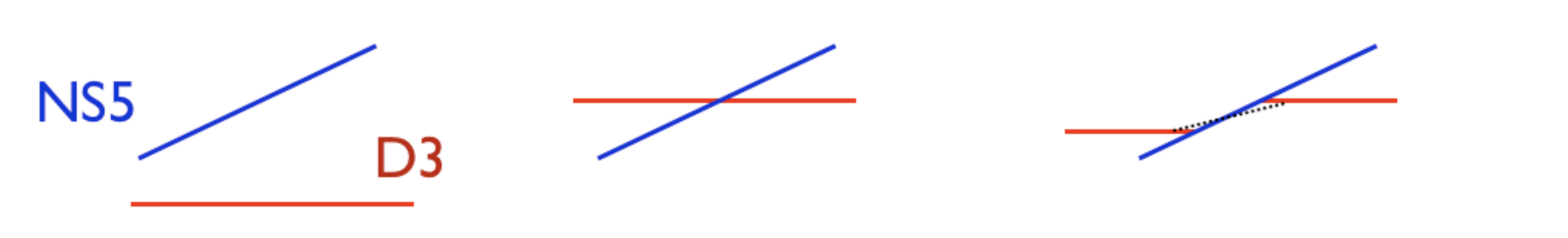} 
   \caption{Bifundamental Multiplet}
   \label{Fig:D3NS5}
   \end{center}
\end{figure}

It is clear from the leftmost diagram of Figure \ref{Fig:D3NS5} that there appears to be a single gauge group on the D3-brane, so one might ask how this fits with our description of two gauge groups and a bifundamental multiplet.
As we are about to see in the next section, the D-flatness conditions state that the expectation value of the Higgs field corresponding to the D3-brane position transverse to the NS5-brane equals the bilinear combination of the bifundamental multiplet $B.$  As a result the Higgs breaks the product of the left and right gauge groups to a subgroup of the skew-diagonal that leaves the bifundamental field invariant.

\section{Impurity Theory and Mirror Symmetry}
\subsection{Theory with Impurities and its D-flatness Conditions}\label{Sec:GaugeTheory}
Supersymmetric gauge theories with impurities of codimension one and two were studied in \cite{Kapustin:1998pb} via the T-duality. We adopt these results to our case in this section. We should mention here the superfield formulations of impurity theories in ${\cal N}=1, d=4$ \cite{DeWolfe:2001pq} and ${\cal N}=2, d=3$ \cite{Erdmenger:2002ex} superspace that can be adopted in our context by introducing defects with bifundamental superfields. Following the approach of \cite{Kapustin:1998pb} here we work in components.   The bosonic matter content in the bulk is 
\begin{itemize}
\item the vector multiplet containing the gauge fields $A_M$ with $M=0,1,2,$ and three Higgs fields $Y^i,$
\item the adjoint hypermultiplet containing two complex Higgs fields $H_1$ and $H_2.$
\end{itemize}
The degrees of freedom localized at the inhomogeneities in the interior at $x_6=\lambda_p$ are two complex fundamental scalars $f_{1p}$ and $f_{2p}.$  The main difference from \cite{Kapustin:1998pb} is that we introduce bifundamental hypermultiplet $B$.  This multiplet contains two scalar fields $B_1$ and $B_2$ transforming as bifundamentals with respect to the gauge groups acting on the left $(x_6=0)$ and the right $(x_6=l)$ ends of the interval.   Augmenting the Lagrangian of \cite{Kapustin:1998pb} with these bifundamental fields we obtain the following component form of the bosonic field Lagrangian $L=L_1+L_2$ of the effective theory on the D3-branes.
The bulk Lagrangian $L_1$ (with the index ranges $M=0,1,2$ and $\mu, \nu=0,1,2,6$) is given by
\begin{align}
L_1=\frac{1}{l} \int d^3x_M dx_6\Bigg\{&
\frac{1}{2} F_{\mu\nu}^2+\frac{1}{2} \big(D_\mu Y^i\big)^2+\frac{1}{2}|[D_\mu H^j]|^2\nonumber\\
&-\frac{1}{2}\sum_{i<j}|[Y^i, Y^j]|^2-\sum_{ij}|[Y^i, H^j]|^2\Bigg\},
\end{align}
and the inhomogeneity Lagrangian
\begin{align}
L_2=&\frac{1}{l } \int d^3x_M dx_6
\Bigg\{
l  \sum_p \delta(x_6-\lambda_p)
		\left(\big|D_M f^p\big|^2-\big|Y^i f^p\big|^2\right)\nonumber\\
&
\qquad \qquad \qquad + \delta(x_6)\left(\big|D_M B\big|^2- \Big|Y^j\vert_{x_6=0+} B-B Y^j\vert_{x_6=l-}\Big|^2\right)
\nonumber\\
&+
\left.
\frac{1}{2}|{\cal D}|^2+{\rm Tr} i {\cal D}_\beta^\alpha\Big([H_\alpha, H^{\dagger\beta}]+
l \sum_p\delta(x_6-\lambda_p) f^p_\alpha\otimes f^{\dagger p\beta}+\right.\nonumber\\
& \qquad \qquad \qquad \qquad - \delta(x_6) B^{\dagger\beta}\otimes B_\alpha+  \delta(x_6-l ) B_\alpha\otimes B^{\dagger\beta} \Big)
\Bigg\},
\end{align}
where ${\rm Im} H_1$ is understood to be a covariant derivative $D_6$ in the sixth direction (in other words $H_1=D_6+T_3=\frac{\partial}{\partial x_6}-i T_0+T_3$),   the auxiliary field 
${\cal D}_\beta^\alpha={\cal D}^j(\sigma_j)_\beta^\alpha,$ and $D_M B$ is the covariant derivative naturally acting on the bifundamenal, e.g. $D_M B_1=\partial_M B_1-i (A_M\vert_{x_6=0+}) B_1+i B_1 (A_M\vert_{x_6=l-}).$

In order to make contact with our notation we let $ \sqrt{l} f_p=\left(\begin{array}{c}
J^\dagger_p\\ I_p\end{array}\right), B=\left(\begin{array}{c}
B_{01}^\dagger\\ B_{10}\end{array}\right),$
 and $H_2=T_1+i T_2,$ then the D-flatness conditions that are easily read off from the above Lagrangian become:
\begin{align}
{d T_1\over dx_6}-&i[T_0,T_1]+i [T_2,T_3]=-{1 \over 2}\bigg\{\sum_{p=1}^k \delta(s-\lambda_p)
\left(IJ-J^\dagger I^\dagger\right)-\nonumber\\
&-\delta(s)(B_{01}B_{10}- B_{10}^\dagger B_{01}^\dagger)+\delta(s-l)( B_{10}^\dagger B_{01}-B_{01}^\dagger B_{10})\bigg\},\\
{d T_2\over dx_6}-&i[T_0,T_2]+i [T_3,T_1]= i {1\over 2}\bigg\{\sum_{p=1}^k
\delta(s-\lambda_p)
\left(I J+J^\dagger I^\dagger \right)- \nonumber\\
&-\delta(s)(B_{01}B_{10}+ B_{10}^\dagger B_{01}^\dagger)+\delta(s-l)( i B_{10}B_{01}+ B_{01}^\dagger B_{10}^\dagger)\bigg\},\\
{d T_3\over dx_6}-&i [T_0,T_3]+i [T_1,T_2]=-{1 \over 2}\bigg\{\sum_{p=1}^k
\delta(s-\lambda_p)
\left(J^\dagger J- I I^\dagger  \right)- \nonumber\\
&-\delta(s)(B_{01}B_{01}^\dagger- B_{10}^\dagger B_{10}^\dagger)+\delta(s-l)( B_{01}^\dagger B_{01}+B_{10} B_{10}^\dagger)\bigg\}.
\end{align}
These reproduce exactly the Nahm equations (\ref{Nahm}), as well as the boundary conditions  (\ref{Eq:jumps1},\ref{Eq:jumps2}) and the matching conditions (\ref{Eq:jumps3},\ref{Eq:jumps4}).

\subsection{Mirror Symmetry and Bow Reciprocity}
The brane configuration identified in Section \ref{Sec:Branes} allows for an S-dual description.  S-duality leads to an analogous configuration with NS five-branes in place of D5-branes and vice versa.   From the point of view of the gauge theory on the D3-brane identified in Section \ref{Sec:GaugeTheory}, this duality is the Montonen-Olive electric-magnetic duality.  The brane picture leads us to conclude that this duality effectively interchanges the two types of inhomogeneities in our theory, leading to a reciprocity among bows.  The reciprocity rule is represented in  Figure \ref{Fig:Reciprocity}. 
\begin{figure}[h]
   \includegraphics[width=6cm]{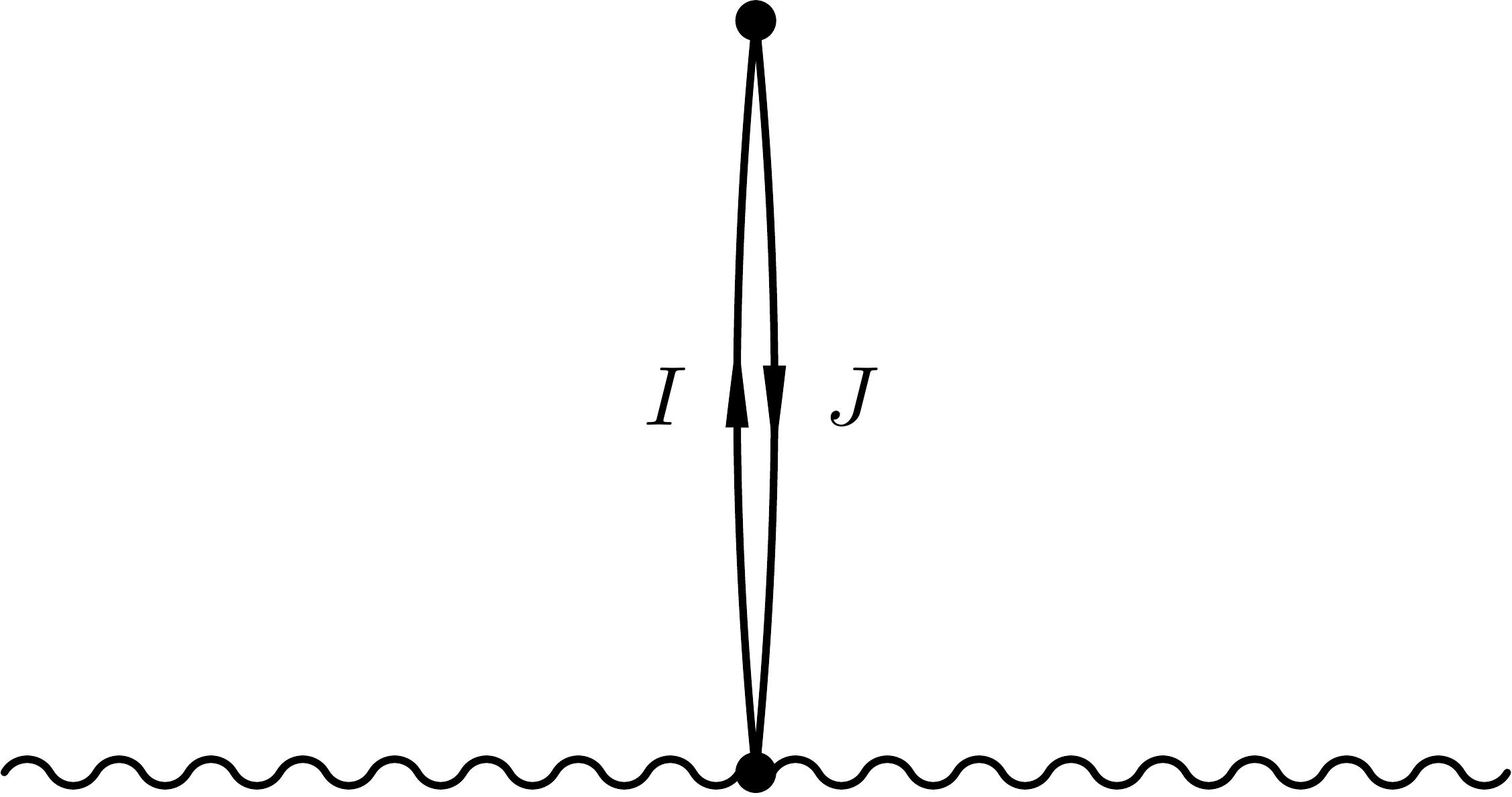} 
   
   \vspace{-2cm}   \hspace{6.5cm}
   \includegraphics[width=0.7cm]{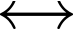}
   
   \vspace{-1.8cm}\hspace{7.5cm}
   \includegraphics[width=6cm]{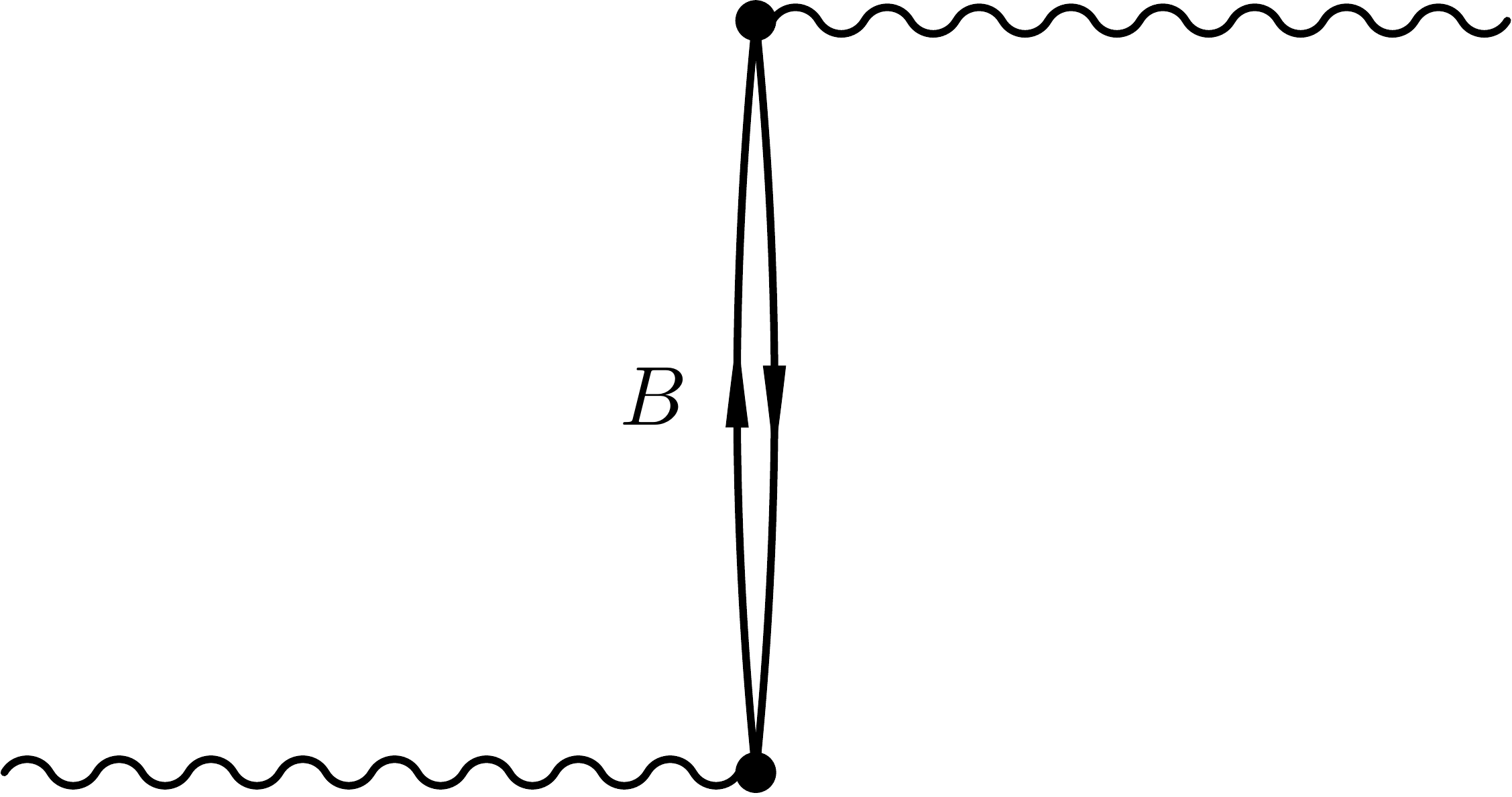}
   \caption{Reciprocity Rule}
   \label{Fig:Reciprocity}
\end{figure}
 It amounts to interchanging the fundamental and bifundamental multiplets and splitting and rejoining the Nahm data intervals accordingly. 
 
If we are to consider a theory with impurities specified by the diagram on the right in Figure~\ref{Fig:Pair}, it is a theory with $SU(N)$ gauge group with three inhomogeneity hyperplanes: two with fundamental and one with bifundamental hypermultiplets.
Two of the branches of vacua of this theory are described by the two bow diagrams of Figure \ref{Fig:Pair}. The Coulomb branch of the theory is isometric to the moduli space of $N$ $U(1)$ instantons on a two-centered Taub-NUT (also called $A_1$ ALF space)\footnote{At first sight this moduli space might appear to be empty, however, in the presence of noncommutativity, indicated here by the difference of the wavy interval lengths, this space is not trivial.}, while the Higgs branch is given by the moduli space of $N$ $U(2)$ instantons on a Taub-NUT space ($A_0$ ALF).  As usual, S-duality maps the above impurity theory to another $U(N)$ impurity theory with two bifundamental and one fundamental matter multiplets.

To give an example, we apply this rule to the bow diagram of $SU(2)$ instantons on the Taub-NUT space.  The reciprocal pair of diagrams is presented in Figure \ref{Fig:Pair}.
\begin{figure}[h]
   \centering
   \includegraphics[width=6cm]{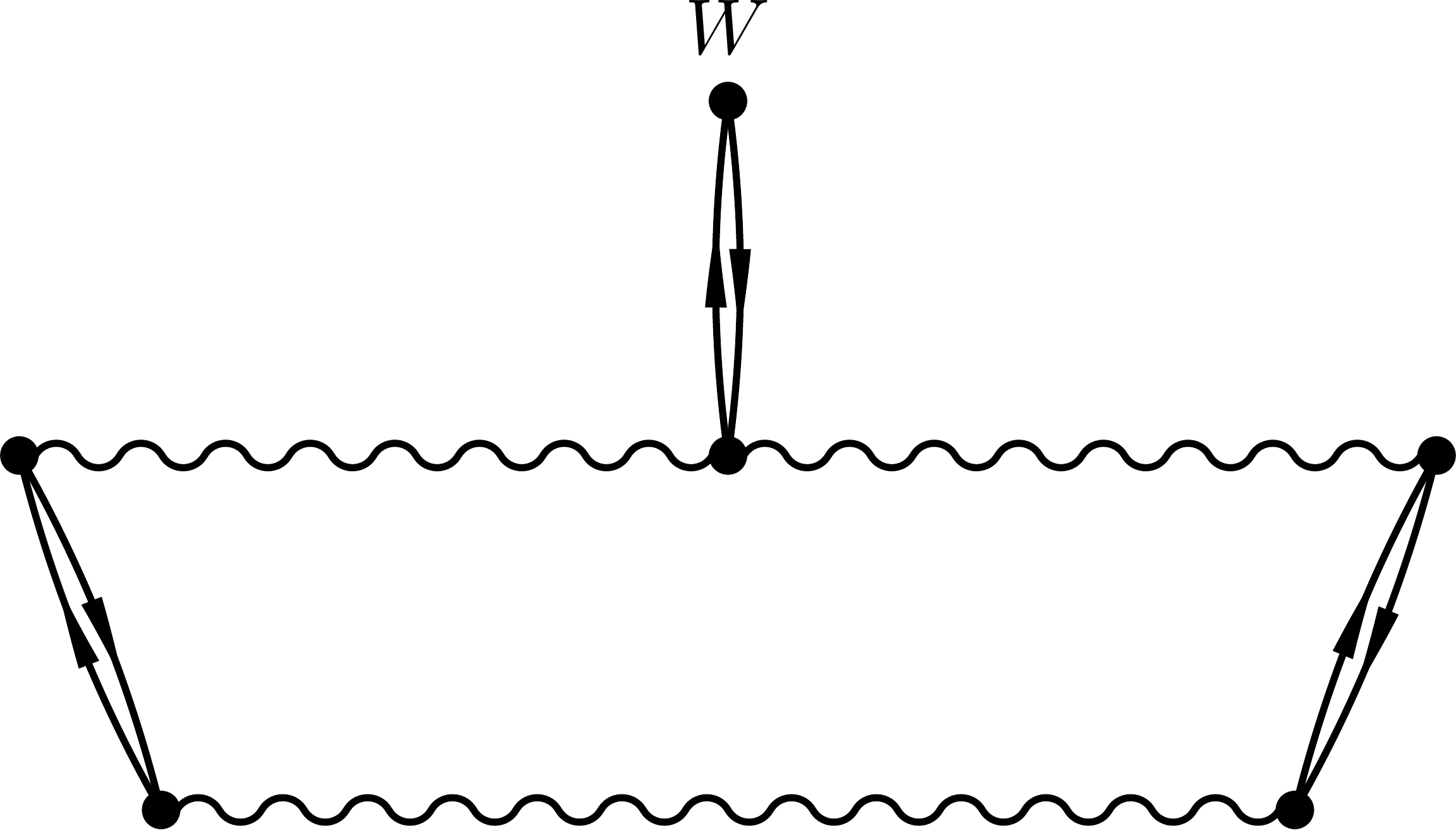} \ \ \ \ \ \ \ 
    \includegraphics[width=5cm]{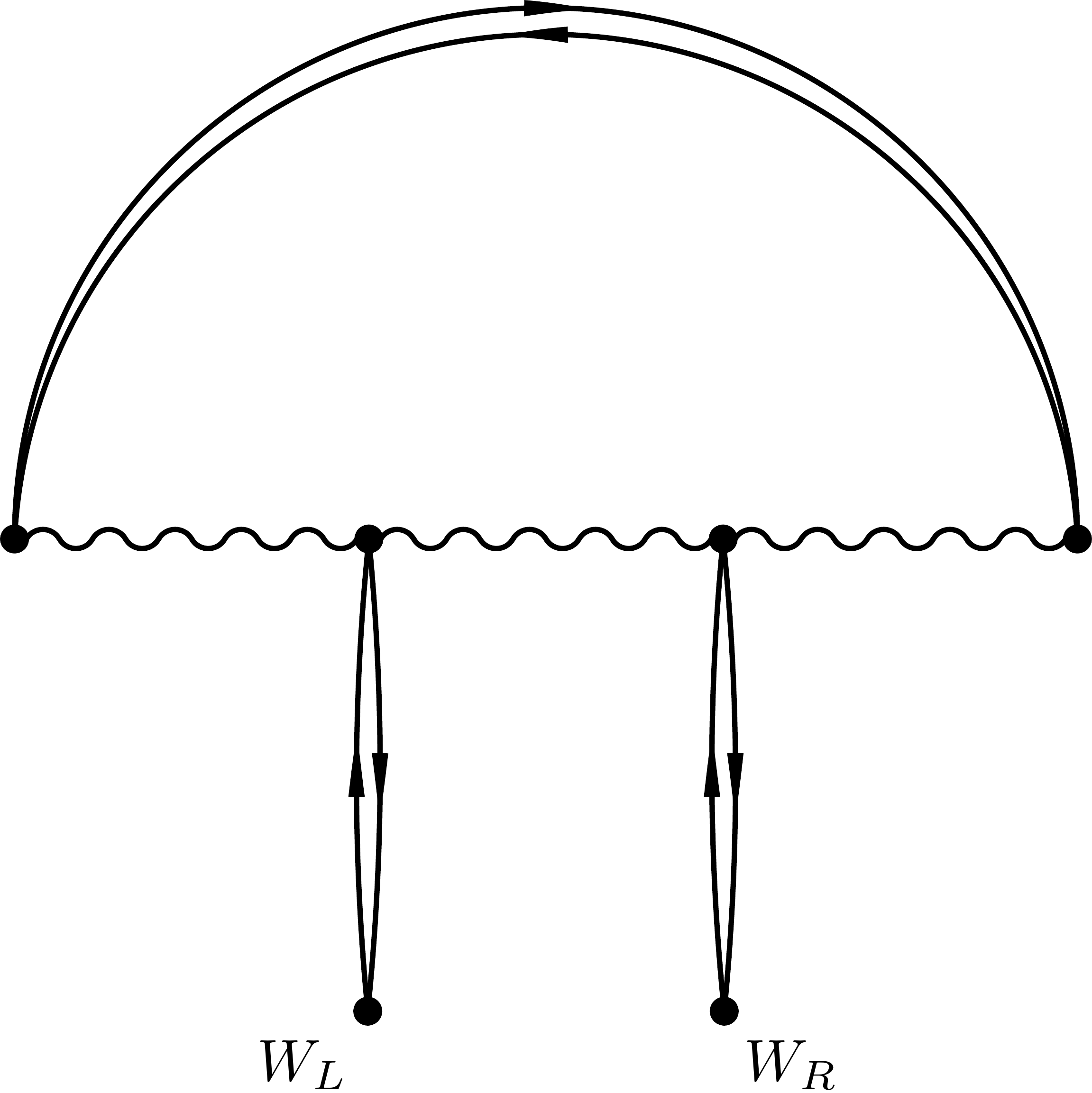}   \caption{A bow pair}
   \label{Fig:Pair}
\end{figure}
The resulting bow diagram describes Abelian self-dual connections on a two-centered Taub-NUT space.  

Let us emphasize that the moduli spaces of a bow and its reciprocal bow are generically different.  What this reciprocity suggests, however, is that the two bows should be considered together and the moduli space of one is viewed as a branch of a larger moduli space.

\section{Moduli Space of One Instanton}
\label{Sec:OneInst}
For the case of a single instanton all the instanton data are Abelian and the Nahm equations imply that all $T_j$'s for $j=1,2,3$ are constant on each interval while $T_0$ can be made constant by a gauge transformation that acts trivially at the boundary.  If the interval has length $d,$ due to a gauge transformation $\exp\left(2\pi i\frac{s-s_0}{d}\right),$ the corresponding $T_0$ is periodically identified with the period $2\pi/d.$  Thus we have three copies of $\R^3\times\S^1$ parameterised by the Nahm data on the three intervals and three copies of $\R^4$ parameterized by $(I_L, J_L), (I_R,J_R),$ and $(B_{01}, B_{10}).$
The remaining gauge group acting on these data is $U(1)^{\times 4}$ with the following action:
\begin{equation}\label{Group}
e^{i\phi_0}\times e^{i\phi_1}\times e^{i\phi_2}\times e^{i\phi_3}: \left(\begin{array}{c}
B_1\\
B_2\\
I_L \\
J_L\\
I_R\\
J_R\\
T_0^L\\
T_0^M\\
T_0^R\\ 
T_j
         \end{array}
	     \right)\mapsto
	     \left(\begin{array}{c}
e^{-i\phi_0} B_1 e^{i\phi_3}\\
e^{-i\phi_3} B_2 e^{i\phi_0}\\
e^{-i\phi_1} I_L \\
J_L e^{i\phi_1} \\
e^{-i\phi_2} I_R \\
J_R e^{i\phi_2}\\
T_0^L+(\phi_1-\phi_0)/d_L\\
T_0^M+(\phi_2-\phi_1)/d_M\\
T_0^R+(\phi_3-\phi_2)/d_R\\
T_j
    	    \end{array}
	     \right)
\end{equation}
It is convenient to introduce quaternionic notation for this data:
\begin{equation}
X_B=\left(\begin{array}{cc}
B_2^\dagger & B_1^\dagger\\
-B_1 & B_2
\end{array}\right),\quad 
X_L=\left(\begin{array}{cc}
I_L^\dagger & -J_L^\dagger\\
J_L & I_L
\end{array}\right),\quad 
X_R=\left(\begin{array}{cc}
I_R^\dagger & -J_R^\dagger\\
J_R & I_R
\end{array}\right).
\end{equation}
Then the complex structures are given by ${\cal I}=-i\sigma_3, {\cal J}=-i\sigma_,$ and ${\cal K}=-i\sigma_2.$
The following computation is close to that of \cite{Gibbons:1996nt}.
For each of the $X_L, X_M,$ and $X_R$ we introduce  coordinates $\Psi_L, \Psi_M,$ and $\Psi_R$ via the following decomposition: $X=Q\exp(-i\sigma_3 \Psi/2),$ with an anti-Hermitian $Q$ and a periodic $\Psi\sim\Psi+4\pi.$  For any such $X,$ the combination $X\sigma_3 X^\dagger$ is traceless and Hermitian, thus it can be written in terms of a real three vector $\vec{R},$ so that $\vec{R}\cdot\vec{\sigma}=X\sigma_3 X^\dagger.$ 
The decomposition of $X$ in terms of $Q$ and $\Psi$ was unique up to a change of the sign of $Q$ assisted by a $2\pi$ shift in $\Psi,$
 thus $X$ is determined uniquely in terms of $\vec{R}$ and $\Psi.$ Moreover, the flat metric
\begin{equation}
 {\mathbb I}_{2\times 2} \cdot ds_X^2 =dX dX^\dagger=\frac{1}{4}\left(\frac{1}{|\vec{R}|} d\vec{R}^2+|\vec{R}|(d\Psi+\Omega)^2  \right)  {\mathbb I}_{2\times 2},
\end{equation}
where $d\Omega=*d\frac{1}{|\vec{R}|}.$
If $X$ transforms as $X\mapsto X\exp(-i\sigma_3\phi)$ (i.e. $\Psi\mapsto\Psi+2\phi$) the corresponding moment map is $\vec{\mu}=\frac{1}{2}\vec{R}.$   On the other hand if the Nahm data transforms as $(T_0, \vec{T}_j)\mapsto(T_0+\phi, \vec{T}_j)$ the moment map is $\vec{\mu}=\vec{T}.$
With this in mind, we find the following moment maps for the $U(1)^{\times 4}$ action of Eq. (\ref{Group}):
\begin{align}
\vec{\mu}_0&=-\frac{1}{2}\vec{R}_B-\vec{T}_L, & \vec{\mu}_1&=\vec{T}_L-\vec{T}_M-\frac{1}{2}\vec{R}_L,\\
\vec{\mu}_3&=-\vec{T}_R+\frac{1}{2}\vec{R}_B&
\vec{\mu}_2&=\vec{T}_M-\vec{T}_R+\frac{1}{2}\vec{R}_R.
\end{align}
The following invariants of the gauge transformations (\ref{Group})
\begin{align}
\theta&=\Psi_B-2d_L T_0^L-\Psi_L+\frac{d_R}{d_L}(\Psi_R+2d_R T_0^R),\\
\alpha&=\Psi_L+\Psi_R-2 d_M T_0^M,
\end{align}
provide two periodic coordinates of period $4\pi$ on the moduli space. 
In our case $d_L=d_R=d/2, l=d_L+d_M+d_R, d_M=2\lambda.$ 
Imposing the vanishing of the moment maps above we have 
\begin{equation}
\vec{T}_L=\vec{T}_R=-\frac{1}{2}\vec{R}_B=\vec{R_1},\quad \vec{T}_M=\vec{R}_2,\quad
 \vec{r}=\frac{1}{2}\vec{R}_L=\frac{1}{2}\vec{R}_R=\vec{R}_1-\vec{R}_2
\end{equation}
\begin{align}
ds^2=&\left(l+\frac{1}{2 R_1}\right) d\vec{R}_1^2-4\lambda d\vec{R}_1 d\vec{r}++\left(2\lambda+\frac{1}{r}\right) d\vec{r}^2\\
&+\frac{\left(\frac{1}{2}d\theta-\frac{1}{4}\omega_{R_1}\right)^2}{l-2\lambda+1/r+1/(2 R_1)}
+\frac{1}{4}\frac{\left(d\alpha+\omega_{r}\right)^2}{2\lambda+1/r},
\end{align}
where $\theta\sim\theta+4\pi, \alpha\sim\alpha+4\pi$ and $d\omega_{R_1}=*d(1/R_1), d\omega_{r}=*d(1/r).$
This metric matches that of \cite{deBoer:1996mp}, where the moduli spaces of the corresponding three-dimentional gauge theories were studied.

We interpret this moduli space metric in the following way.  Just as in the case of a caloron, an instanton on the Taub-NUT space has two monopole-like constituents.  These constituents are characterized by their positions in the three-space given by the vectors $\vec{R}_1$ and $\vec{R_2}$ and phases $\theta_1$ and $\theta_2$ respectively.  The metric above is written in terms of the coordinates and the phase of the first constituent $\vec{R}_1$ and $\theta=\theta_1$ and the relative position $\vec{r}=\vec{R}_1-\vec{R}_2$ and the relative phase $\alpha=\theta_2-\theta_1$ of the two constituents.

\section{Conclusions}
We presented Bow Diagram formalism which encodes the data determining instanton configurations on the Taub-NUT and multi-Taub-NUT spaces.  Our discussion here was limited to zero monopole charge and a generic Wilson line at infinity.  We motivated our construction by identifying a corresponding string theory brane configuration and analyzing the theory on D-branes' world-volume.  The resulting impurity theory on a four-dimensional space-time with one compact dimension contains two types of impurities localized on hyperplanes perpendicular to the periodic direction.  We formulate a reciprocity rule that interchanges the two types of impurities. Applying this rule to all impurities in a bow leads to a dual bow.  Supersymmetry conditions defining the moduli space of the quantum gauge theory are exactly the moment maps of the corresponding bow.

The bow formulation allows us to find the moduli space of instantons on the Taub-NUT.  We identify it as a finite hyperk\"ahler quotient and establish its holomorphic equivalence with the moduli space of instantons on $\R^4.$  As an example, we find the metric on the moduli space of a single instanton on the Taub-NUT space.

\section*{Acknowledgments} 
It is our pleasure to thank Tamas Hausel and Juan Maldacena for illuminating discussions. We are grateful to the Institute for Advanced Study, Princeton and Princeton University Mathematics Department for hospitality.

This work is supported by the Science Foundation Ireland Grant No. 06/RFP/MAT050 and by the European Commision FP6 program MRTN-CT-2004-005104.

\end{document}